\documentclass[conference]{IEEEtran}  

\IEEEoverridecommandlockouts                              

\usepackage{amsmath} 
\usepackage{pgf}
\usepackage{subfigure}
\usepackage[export]{adjustbox}
\usepackage{algorithm}
\usepackage{algorithmicx,algpseudocode}
\usepackage[normalem]{ulem}
\usepackage{lipsum}
\usepackage{mathtools}
\usepackage{cuted}
\usepackage{multirow}
\usepackage{graphicx}
\graphicspath{{./figures/}}

\begin{document}
\title{\LARGE \bf
Optimal Placement of Baseband Functions for Energy Harvesting Virtual Small Cells
}

\author{\IEEEauthorblockN{Dagnachew A. Temesgene, Nicola Piovesan, Marco Miozzo, Paolo Dini}\\
    \IEEEauthorblockA{CTTC/CERCA, Av. Carl Friedrich Gauss, 7, 08860, Castelldefels, Barcelona, Spain
    \\\{dtemesgene, npiovesan, mmiozzo, pdini\}@cttc.es}
    }
   

\maketitle
\thispagestyle{empty}
\pagestyle{empty}

\begin{abstract}

Flexible functional split in Cloud Radio Access Network (CRAN) greatly overcomes fronthaul capacity and latency challenges. In such architecture, part of the baseband processing is done locally and the remaining is done remotely in the central cloud. On the other hand, Energy Harvesting (EH) technologies are increasingly adopted due to sustainability and economic advantages. Power consumption due to baseband processing has a huge share in the total power consumption breakdown of smaller base stations. Given that such base stations are powered by EH, in addition to QoS constraints, energy availability also conditions the decision on where to place each baseband function in the system. This work focuses on determining the performance bounds of an optimal placement of baseband functional split option in virtualized small cells that are solely powered by EH. The work applies Dynamic Programming (DP), in particular, Shortest Path search is used to determine the optimal functional split option considering traffic QoS requirements and available energy budget.

\end{abstract}


\section{Introduction}
Cloud Radio Access Network (CRAN) ensures cloudification of RAN by centralized pooling of the Baseband (BB) processing units ~\cite{chinamobile}. The need for high capacity and very low latency fronthaul is the major challenge behind the full realization of CRAN. To alleviate this problem, flexible functional split between local base station (BS) sites and a central  Baseband Unit (BBU) pool is proposed~\cite{scforum}. By executing some baseband processing tasks locally, the tight latency and bandwidth requirement of the fronthaul can be relaxed, while maintaining many of the centralization advantages that CRAN architecture offers. In addition, Heterogeneous CRAN (HCRAN) is proposed as an architecture that includes a presence of High-Power Nodes (HPNs) for control plane functions and coverage \cite{peng2014heterogeneous} to partially alleviate the fronthaul constraint in CRAN.

On the other hand, energy sustainability is one of the key pillars for future mobile network design and operation due to environmental concerns and costs. For this reason, Energy Harvesting (EH) technology is increasingly adopted as a means to ensure sustainability for next generation mobile \mbox{networks \cite{piro2013}}. However, EH comes with its own unique challenge mainly due to unreliable energy sources.  Hence, in Energy Harvesting Base Stations (EHBSs), it is important to properly manage the harvested energy and to ensure proper energy storage provision to avoid energy outage.  

Most of the literature on energy management policy in EHBSs focus on a HetNet architecture with an intelligent switching on/off scheduling of base stations. The authors in~\cite{lee2017} apply a ski-rental framework based online algorithm for optimal switch on/off scheduling of EHBSs.  On the other hand, the authors in~\cite{miozzo2015} apply reinforcement learning to optimize the energy usage.  A similar problem of energy saving optimization in HetNets is studied in~\cite{ameur2016}, where multi-armed bandit is applied to determine optimal cell expansion bias. The authors in~\cite{piovesan2017} apply Dynamic Programming (DP) to determine the optimal switch on and off policy of a HetNet by considering the traffic variation and energy arrival.  

Nevertheless, embedding EH and incorporating flexible functional split options in HCRAN is missing in the literature. The functional splits give insight into considering more operation modes of BSs, in addition to switch on and off. This paper fills this gap by proposing an optimal energy management scheme that incorporates functional split options. The proposed approach is unique by considering more operational modes of virtual small cells as compared to the current literature that focus only on intelligent switching on and off policies. 

The main contributions of the paper are: 
\begin{itemize}
\item Formulating the energy management of virtual EHBSs and a central BBU pool as an offline optimization problem targeting three functional split options, namely MAC/PHY, UpperPHY/LowerPHY and CRAN;
\item Applying a DP algorithm to find the optimal placement of functional split options considering the traffic demand, energy reserve, forecasted energy arrival and QoS constraints. In particular Shortest Path search is applied for solving the optimization problem;
\item Presenting the performance of dynamic placement of functional split options through numerical results with comparison against static configurations. Hence, these results can serve as performance bounds for online optimization approaches.
\end{itemize}

The rest of the paper is organized as follows. Section~\ref{sec:network-scenario} describes the considered network scenario. Section~\ref{sec:system-model} introduces the system model. In Section~\ref{sec:algo}, the optimization problem and the optimal algorithm are described. The results are discussed in Section~\ref{sec:results}. Finally, we draw our conclusions in Section~\ref{sec:conclusions}.
\section{ Network Scenario}
\label{sec:network-scenario}
We consider a RAN as a set of clusters of one Macro Base Sation (MBS) with co-located BBU pool and $N$ virtual Small Cells (vSCs). 
The vSCs provide service to hotspots, whereas mobility and baseline coverage are provided by the MBS. The vSCs are fully powered by EH plus batteries and are endowed with limited computational resource that can be used opportunistically, e.g., when enough energy is available, for part of the baseband signal processing tasks. The MBS with the BBU pool is powered by the electrical grid. The connection between vSCs and MBS is provided by a reconfigurable fronthaul and the BBU pool is capable of performing part of the baseband processing. 

The functional split options that can be applied for the vSCs are given in~\cite{scforum}. Considering the potential centralization gains, we have selected the following functional split options as targets in this paper (shown in Fig. \ref{fig:signalflow}):

\begin{itemize}
\item Standard CRAN -- all the baseband processing is done centrally at the BBU pool;
\item UpperPHY/LowerPHY -- the LowerPHY layer processing is done by the vSC whereas UpperPHY and above is executed by the BBU pool;
\item MAC/PHY -- the whole PHY layer processing takes place at vSCs whereas MAC and above layers are done at the central BBU pool.
\end{itemize}

\section{System Model}
\label{sec:system-model}
A state vector $\boldsymbol{S}_t= [S_t^1,S_t^2,\ldots,S_t^N ]$ represents the possible operative mode of each vSC at time $t$ . Each single element $S_t^i$ is defined as:

\begin{equation}
\boldsymbol{S}_t^i= \begin{cases} 0, & \mbox{if } i^{\mathrm{th}}\mbox{ vSC is switched off} \\ 1, & \mbox{if } i^{\mathrm{th}} \mbox{ vSC is in CRAN split mode} \\ 2, & \mbox{if } i^{\mathrm{th}} \mbox{ vSC is in UpperPHY/LowerPHY mode} \\ 3, & \mbox{if } i^{\mathrm{th}} \mbox{ vSC is in MAC/PHY  mode}\end{cases}
\label{eq:system}
\end{equation}

The energy harvested by each vSC at time $t$ is defined by the vector $\boldsymbol{E}_t= [E_t^1,E_t^2,\ldots,E_t^N ]$ whereas the energy stored by each vSC at time $t$ is defined by the vector $\boldsymbol{B}_t= [B_t^1,B_t^2,\ldots,B_t^N ]$. The traffic load experienced by each vSC is defined by the vector $\boldsymbol{\rho}_t= [\rho_t^1,\rho_t^2,\ldots,\rho_t^N ]$.

\subsection{Power model}

\begin{figure}[b]
\centering
\includegraphics[scale = 0.12]{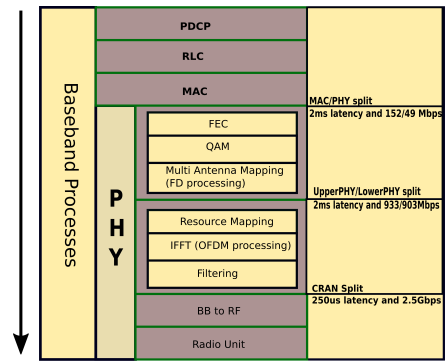}
\caption{BBU functions with the considered functional splits and the relevant fronthaul latency and bandwidth requirements (estimated based on \cite{scforum})} 
\label{fig:signalflow}
\end{figure}

The power consumption of each split option is estimated based on the model introduced in \cite{desset2012}, which is a general flexible power model of LTE base stations and provides the power consumption in Giga Operation Per Second (GOPS). Technology dependent GOPS to Watt conversion factor is applied to determine the power consumption in Watts. 
In this paper, we have mapped the various baseband processing tasks of the functional split options to their power requirement estimations. 
The main baseband tasks associated with the split options are shown in Fig. \ref{fig:signalflow}. 

The total base station power consumption is given by:

\begin{equation}
P_{\mathrm{BS}}=P_{\mathrm{BB}}+P_{\mathrm{RF}}+P_{\mathrm{PA}} + P_{\mathrm{overhead}}
\label{eq:Ptotal}
\end{equation}
where $P_{\mathrm{BB}}$ is the power consumption due to baseband processing, $P_{\mathrm{RF}}$ is the power consumption due to RF, $P_{\mathrm{PA}}$ is the power consumption by the power amplifier and $P_{\mathrm{overhead}}$ is the overhead power consumption, e.g., cooling system. 

The baseband power consumption, $P_{BB}$, is generally computed as:
\begin{equation}
P_{\mathrm{BB}}  =P_{\mathrm{BB1}}+ P_{\mathrm{BB2}} 
\label{eq:PBB}
\end{equation}

More in detail, $P_{\mathrm{BB1}}$ is given by:
\begin{equation}
P_{\mathrm{BB1}}  =[P_{\mathrm{CPU}}+ P_{\mathrm{OFDM}}+ P_{\mathrm{filter}} ]
\label{eq:PBB1}
\end{equation}
where $P_{\mathrm{CPU}}$ is the idle mode power consumption, $P_{\mathrm{OFDM}}$ is the power consumption due to OFDM processes and $P_{\mathrm{filter}}$ is the power consumption due to filtering.
In addition, $P_{\mathrm{BB2}}$, is given by:
\begin{equation}
P_{\mathrm{BB2}}  = [P_{\mathrm{FD}}+ P_{\mathrm{FEC}} ]
\label{eq:PBB2}
\end{equation}
where  $P_{\mathrm{FD}}$ is the frequency domain processing power consumption and $P_{\mathrm{FEC}}$ is the power consumption due to FEC processes.
Estimating these power consumption values mainly depends on bandwidth, number of antennas and the load fraction. In particular, $P_{\mathrm{FD}}$  and $P_{\mathrm{FEC}}$  are dependent on the traffic load. The power dependence on these factors can be both linear and exponential~\cite{desset2012}. 

The baseband power consumption of the vSC depends on the adopted functional split option, in particular it is given as:
\begin{equation}
P_\mathrm{BB}^{\mathrm{vSC}}= \begin{cases}  0, & \mbox{if vSC is in CRAN}  \\ P_\mathrm{BB1}, & \mbox{if vSC is in UpperPHY/LowerPHY} \\ P_\mathrm{BB1}+P_\mathrm{BB2}, & \mbox{if vSC is in MAC/PHY} \end{cases}
\label{eq:system_BBvSC}
\end{equation}
The power consumption of the MBS is determined by (\ref{eq:Ptotal}). Additional MBS power consumption is considered based on the functional split option of the vSCs in the same mobile cluster (e.g., additional $P_\mathrm{BB2}$ for each vSCs in UpperPHY/LowerPHY split).

\section{Optimal Solution}
\label{sec:algo}

\subsection{Problem Statement}
The intelligent energy management decision is a sequential process that selects the optimal configuration for each vSCs based on the traffic demand, the energy reserve, the forecasted energy arrival and QoS constraints. The objective is to minimize the power drained from the grid which, in turn, is equivalent to MBS power consumption, and avoid system outage. We define system outage as the event to not satisfy the traffic demand due to battery energy depletion or wrong configuration decisions. These wrong configurations may overload the MBS radio access with the traffic of the handed over UEs of the switched off vSCs.

 The process evolves in cycles along with the traffic demand and the energy arrival variations. At each cycle $t$, the task of the centralized controller is to select the optimal operative mode of each vSC among the four options given by (\ref{eq:system}).
The minimization of the MBS power consumption is modeled as a DP optimization problem and it is given by:
\begin{equation}
\begin{split}
\min_{\{\boldsymbol{S}_t\}_{t=1,\dots,K}}\; &\sum_{t=1}^K f(\boldsymbol{S}_t,t) \\
B_t^{(i)}&>B_{\mathrm{th}} \; \forall i .\\
\\
\end{split}
\label{eq:optimization}
\end{equation}
where $B_{\mathrm{th}}$ is the battery threshold level adopted to prevent damages to the storage devices and $K$ is the time horizon or the number of times the energy control is applied.
The cost function is defined as:
\begin{equation}
f(\boldsymbol{S}_t,t)=w_1\cdot \mathrm{P_m}(\boldsymbol{S}_t,t)+ w_2\cdot {\mathrm{D}}(\boldsymbol{S}_t,t)
\label{eq:cost}
\end{equation}
where $\mathrm{P_m}(\boldsymbol{S}_t,t)$ and  $\mathrm{D}(\boldsymbol{S}_t,t)$ are respectively the grid power consumption and the traffic drop rate of the system given the operative modes of the vSCs and the time step. The weights $w_1$ and $w_2$  provide flexibility in the cost function to emphasize one part of the cost over the other. They must always sum to 1, that is, $w_1+ w_2=1$.

\subsection{Graphical Representation}
\begin{figure}
\centering
\includegraphics[scale = 0.32]{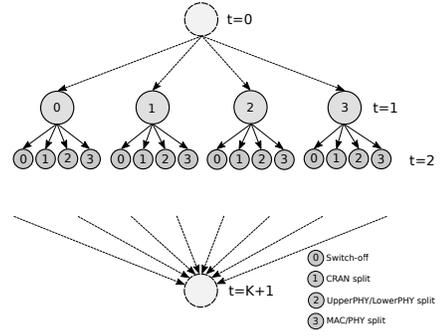}
\caption{Graphical representation of the functional split configuration decision process for the case of one vSC} 
\label{fig:problem}
\end{figure}

The problem of finding the optimal operating modes is represented as a graph. In the graph, a single node $(N_t^i)$ represents a possible combination of  operating modes of the vSCs. These combinations result in different grid power consumptions, system drop rates and energy storage levels of vSCs.
In Fig.~\ref{fig:problem}, an example of graph for a system with a single vSC is depicted. At first time step $(t = 1)$, the vSC can be in one of the four possible operating modes: switch off, CRAN split mode, UpperPHY/LowerPHY split mode and MAC/PHY split mode. Moving one time step ahead, the energy harvesting and traffic demands are also evolving. Hence, each node $(N_t^i)$  generates four possible child nodes corresponding to the four possible operating modes at cycle $t+1$, $(N_\mathrm{t+1}^j), j=1,...,4$. 
The number of such possible combinations keeps on evolving until reaching the time horizon $K$, leading to the maximum number of possible paths at time instant $K$.

At cycle $t+1$, the battery level corresponding to the child nodes is calculated based on:
\begin{equation}
\boldsymbol{B}_{\mathrm{t+1}}  = \min ( \boldsymbol{B}_{\mathrm{t}}+  \boldsymbol{E}_{\mathrm{t}}, B_{\mathrm{cap}}) -  \mathbf{P}_{\mathrm{vSC} }(\mathbf{S}_t)\cdot \Delta_t
\label{eq:battery}
\end{equation}
where $B_{\mathrm{cap}}$ is the maximum capacity of the battery in $\mathrm{kWh}$, $\Delta_t$ is the time between two consecutive cycles, $\mathbf{P}_\mathrm{vSC}(\mathbf{S}_t)$ is a vector representing the power consumption of the vSCs depending on their operative modes.
The cost function (\ref{eq:cost}) is used to compute the cost associated to each arc connecting two nodes.
Two artificial nodes have been added at time step $t = 0$ and $t = K + 1$, to have a single initial node and a single terminal node. The cost associated to the arcs connecting the artificial nodes are set to zero. 
The cost associated to each arc of the graph can be interpreted as its length. Hence, the optimization problem in ~(\ref{eq:optimization}) is equivalent to finding the shortest path from the initial node at time $t=0$ to the terminal node  at time $t= K+1$. 

\subsection{Shortest Path Search}

We consider the Label Correcting Algorithm~\cite{bertsekas1995dynamic} for finding the shortest path. The exploration of the graph is done in a depth-first approach by sequentially discovering shorter paths from the starting node to the intermediate nodes until reaching the destination node. We define the variable $d_i$, called {\it label of i}, as the length of the shortest path to the node $i$, OPEN as the list of nodes to be explored and UPPER as the last found minimum-length path. Throughout the exploration process, the length of the shorter path found so far is maintained in $d_i$. If a new path is found with shorter length to $i$, the algorithm considers whether the labels $d_j$ of the child nodes $j$ can be corrected by setting $d_j$ to $d_i+a_{ij}$, where $a_{ij}$ is the $arc(i,j)$. The nodes that are candidates to be included in the shortest path are maintained in the list OPEN. Nodes that result in a path length longer than UPPER or those that cannot satisfy the battery and system constraints are excluded from this candidate list. The steps of the algorithm are shown in Algorithm~\ref{alg:algorithm}. This exploration policy is relatively faster and requires lower memory by avoiding to explore the whole \mbox{graph \cite{bertsekas1995dynamic}.} This is especially advantageous for a tree-like problem such as the one tackled in this work. 
\begin{algorithm}[H]
\caption{Shortest Path Search Algorithm}
\begin{algorithmic} 
\State initialize OPEN with possible states at time $t$
\While {OPEN is not empty}
\State remove a node $N_t^i$
\State compute $\boldsymbol{B}_{t+1,j}, j=1,..,4^{N}$, for all $\boldsymbol{S}_{t+1}$ using (\ref{eq:battery})
\For {each node $N_{t+1}^j$ child of $N_t^i$}
\State $a_{ij}=f(\boldsymbol{S}_{t+1,j},t+1)$
\If{$d_i+a_{ij}<\min\{d_j,\mathrm{UPPER}\}$ \\
\begin{center}{\bf and} $\boldsymbol{B}_{t+1,j}>B_{\mathrm{th}}$ }
\end{center}
\State $d_j \leftarrow d_i+a_{ij}$
\State set $N_t^i$ parent of $N_{t+1}^j$
\If {$t\neq K$}
\State place $N_{t+1}^j$ in OPEN (if not already)
\Else
\State $\mathrm{UPPER}=d_i+a_{ij}$
\EndIf
\EndIf
\EndFor
\EndWhile
\end{algorithmic}
\label{alg:algorithm}
\end{algorithm}
\section{Results and Discussion}
\label{sec:results}

\subsection{Simulation Scenario}

We consider an area of $1\times1$ km$^2$ covered by a MBS with co-located BBU pool placed at the center and connected to the electrical grid. Capacity enhancement is provided by 3 vSCs equipped with a solar panel and a battery. A time horizon of $21$ hours (i.e., $K = 21$ ) is selected as it represents a reasonable balance between algorithm performance and complexity as described in \cite{piovesan2017}.
User activities are categorized based on~\cite{auer2010} as heavy users with an activity of $900$ MB/hr and ordinary users with an activity of $112.5$ MB/hr. 
Moreover, we adopt the traffic profiles described in~\cite{xu2017}, in particular the residential and office ones. 
The solar energy traces are generated using the SolarStat tool~\cite{miozzo2014} for the city of Los Angeles and simulations are carried out for the months of January (representing the worst energy harvesting month) and July (representing the best harvesting month). 

The reference vSC power consumption values for our scenario are $P_\mathrm{RF}=2.6$ W and $P_\mathrm{PA}=71.4$ W. For $P_{BB}$, we consider $200$ GOPS, $160$ GOPS and $80$ GOPS for $P_{\mathrm{CPU}}$, $P_{\mathrm{filter}}$ and  $P_{\mathrm{OFDM}}$ respectively. Moreover, the reference load dependent power consumption values are  $30$ GOPS, $10$ GOPS and $20$ GOPS for linear component of $P_{\mathrm{FD}}$, non-linear component of $P_{\mathrm{FD}}$ and $P_{\mathrm{FEC}}$ respectively. 
As for the MBS, we consider $P_\mathrm{RF}=9.18$ W and $P_\mathrm{PA}=1100$ W. The baseband power consumption is 630 GOPS and 215 GOPS for the static ($P_\mathrm{CPU}+P_\mathrm{OFDM}+P_\mathrm{filter}$) and load dependent components ($P_\mathrm{FD}+P_\mathrm{FEC}$), respectively. The power consumption overhead ($P_{\mathrm{overhead}}$) is of 0 and 10\% of the total power of the rest of the base station for the case of vSC and MBS, respectively. Other simulation parameters are given in Table~\ref{tab:table1}.

\begin{table}[ht]
  \centering
  \caption{Simulation Parameters.}
  \begin{tabular}{l c}
   Parameter & Value\\
   \hline \hline
   Solar panel size ($\mathrm{m}^2$) & 4.48 \\
   Solar panel efficiency (\%) & 20\\
   Transmission power of macro cell (dBm) & 43\\
   Transmission power of vSC (dBm) & 38\\
   Bandwidth (MHz) & 5\\
   Antenna & 2x2\\
   Battery capacity (kWh) & 2\\
   Battery threshold (\%) & 20\\
   GOPS to Watt conversion factor & 8\\
  \hline
  \end{tabular}
\label{tab:table1}
\end{table}

\subsection{Optimal Functional Split Configurations}
\begin{figure*}
\centering
\includegraphics[scale=0.8]{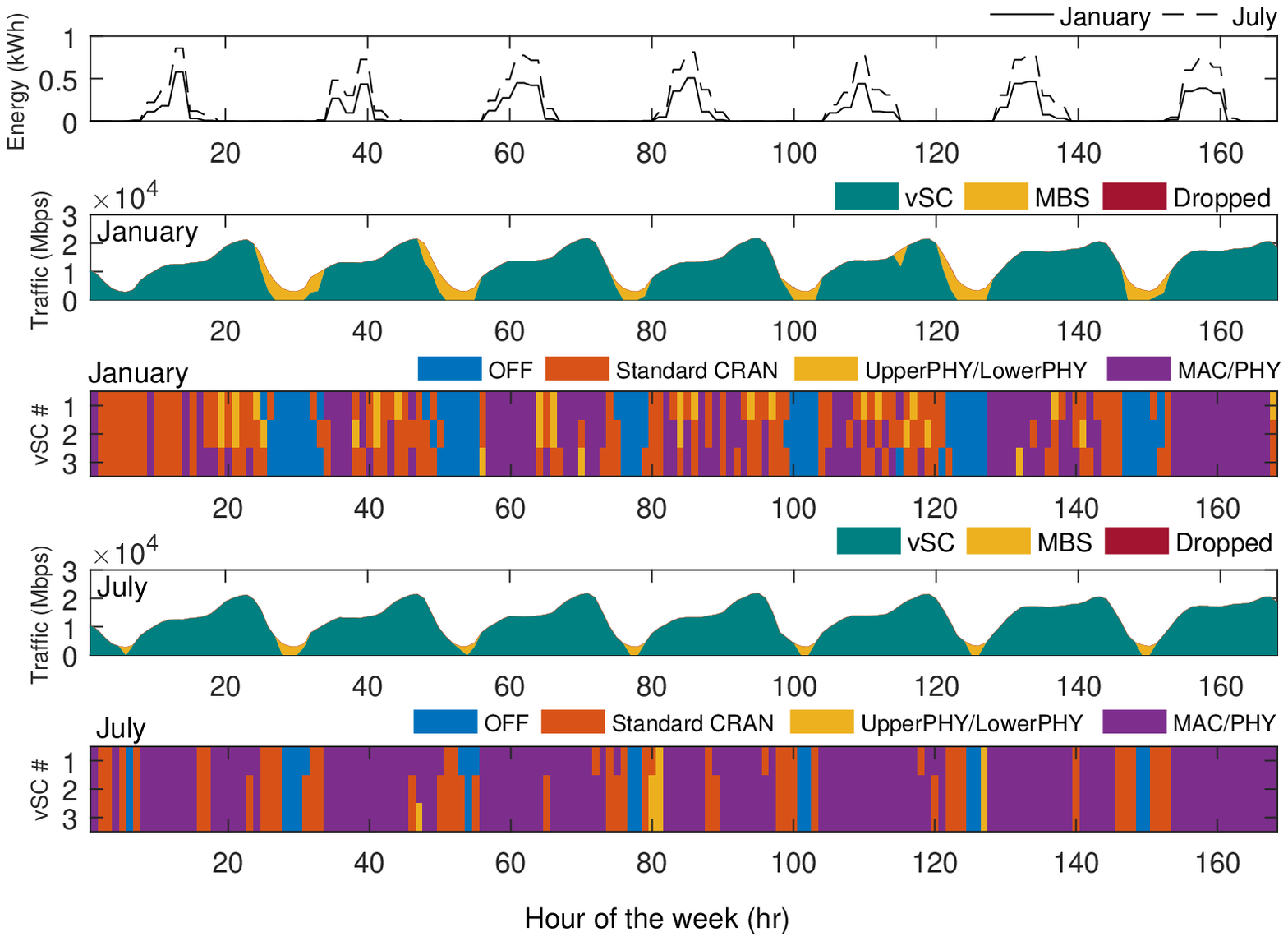}
\caption{Optimal functional split placement results in a residential area scenario for a week of January and July  (hour 0 to hour 168; Monday from 0 - 23 hr). The traces show the amount of harvested energy, the amount of mobile traffic handled by vSCs and MBS and operative mode of each vSCs for both January and July weeks.}
\label{fig:result_r}
\end{figure*}
\begin{figure*}
\centering
\includegraphics[scale=0.8]{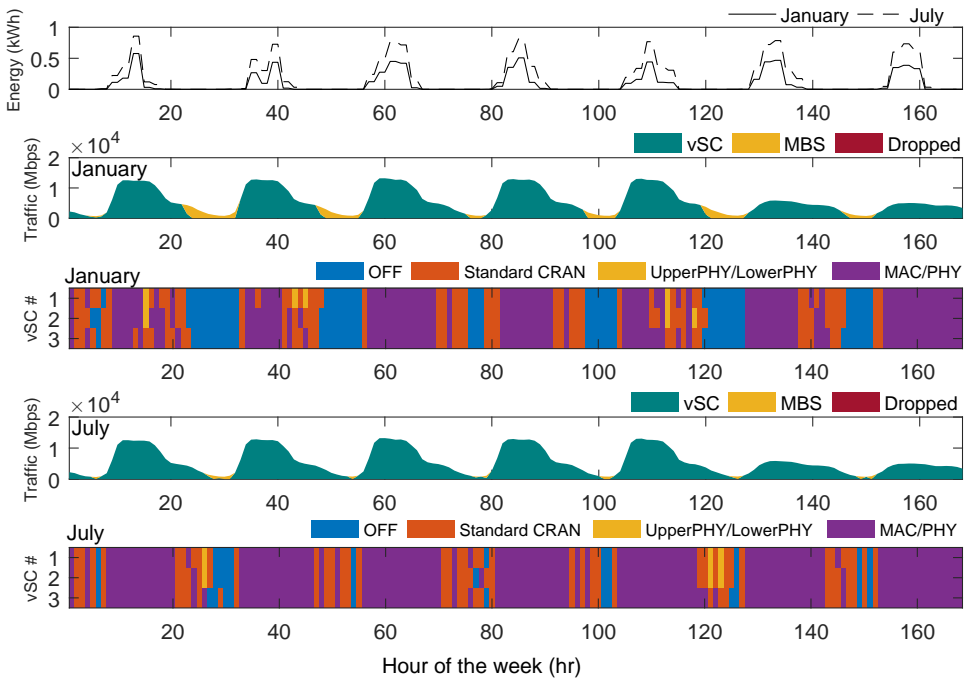}
\caption{Optimal functional split placement results in an office area scenario for a week of January and July (hour 0 to hour 168; Monday from 0 - 23 hr). The traces show the amount of harvested energy, the amount of mobile traffic handled by vSCs and MBS and operative mode of each vSCs for both January and July weeks.}
\label{fig:result_o}
\end{figure*}

The result of optimal functional split placement decisions for a scenario involving 3 vSCs with $90$ users per vSC in a residential area is shown in Fig.~\ref{fig:result_r}. Heavy users ratio of 50\% is considered. The policy is able to dynamically decide the placement of the baseband functions in accordance with the available energy, forecasted harvested energy and traffic demands. Hence the result shows that, most of the user traffic is handled by the vSCs with no outage. Moreover, for a January week, the CRAN and MAC/PHY are the most chosen split options during daytime and peak traffic periods whereas switching off occurs during very low traffic hours. The average CRAN and MAC/PHY selection rate is 37\% each and switching off rate is averaged at 20\%. The results of the simulation for a week of July shows that MAC/PHY split mode is the most selected option. Average MAC/PHY selection rate is 67\% whereas switching off rate is averaged at 8\%. This confirms that due to high energy availability, the vSCs are performing most of the baseband processes by themselves which, in turn, further reduces the grid energy consumed by the MBS.

The result of optimal functional split placement for a scenario of 3 vSCs deployed in an office area with $90$ users per vSC is shown in Fig.~\ref{fig:result_o}. Heavy users ratio of 50\% is considered. An office area traffic profile is characterized by relatively lower traffic peak both in weekdays and weekends. In addition, the peak traffic hours are different than the residential traffic profile. The result shows that the dynamic placement of the baseband functions enable the vSCs to offload the MBS for most of the users traffic without any drop. 
In addition, MAC/PHY and CRAN operative modes are the most selected options in January during peak traffic periods.  The average MAC/PHY and CRAN selection rate is 47\% and 28\% respectively. Average switch off rate is at 23\% and occurs during very low traffic periods, i.e. during night. In July, MAC/PHY is the most selected operative mode  with an average selection rate of 68\% and occurring during peak traffic periods, whereas, switch off occurs during very low traffic hours with average rate of 7.3\%. This is due to the higher energy income in July. The high selection rate of MAC/PHY split results in a further reduction of grid energy consumption, since most of the baseband processes are performed locally by vSCs. 

\subsection{Comparison with static policies}
This subsection provides a comparative analysis of the proposed optimal solution with static functional split policies. The vSCs are kept in the same split mode as long as the battery level is above the threshold, otherwise it is switched off.
The results of these static policies for a scenario with 3 vSCs deployed in a residential area with $90$ users each and a heavy user ratio of $50$\% for a week of January and July are shown in Table ~\ref{tab:table3}. The optimal policy clearly outperforms the static policies regarding traffic drop rate. Moreover, the CRAN policy shows smaller outage against both UpperPHY/LowerPHY and MAC/PHY policies. This is mainly due to the low energy consumption of the vSCs in CRAN mode since they do not perform any baseband operation. Both UpperPHY/LowerPHY and MAC/PHY polices experience high drop rates for the case of January. The optimal policy also outperforms the static policies in terms of grid energy consumption. Saving of up to $24$~KWh and $29$~KWh are achieved respectively in January and July for only a week of operation.

\begin{table}[ht]
  \centering
\caption{Policy Comparisons}
\scalebox{1}{
\begin{tabular}{|c|c|c|c|c|c|}
\hline
\multirow{2}{*}{Policy} & \multicolumn{2}{|p{2cm}|}{\centering Grid energy consumption (KWh)} & 
    \multicolumn{2}{|p{2cm}|}{\centering Average drop rate (\%)} \\
\cline{2-5}
 & January & July & January & July \\
\hline
Optimal & 149.51& 133.23& 0& 0\\

CRAN & 170.01& 162.48& 2.35& 1.5 \\

UpperPHY/LowerPHY & 173.76& 153.64& 16.43& 5\\

MAC/PHY & 173.56& 151.40& 17.10 & 5\\
\hline
\end{tabular}}
\label{tab:table3}
\end{table}

\section{Conclusions}
\label{sec:conclusions}
 In this paper, we have proposed an optimal functional split placement of energy harvesting virtual small cells that relies on central BBU pool for part of baseband processing. Dynamic Programming and more specifically a Shortest Path search algorithm is applied to determine the optimal functional split configurations. In particular, three functional split options namely, CRAN, UpperPHY/LowerPHY and MAC/PHY, have been targeted. Simulation results show that the optimal placement of functional split options results in an improved QoS and  significant energy saving, and hence lower OPEX, with respect to having a static functional split policy. The obtained performance bounds represent an encouraging starting point for online optimization approaches.

\addtolength{\textheight}{-12cm}   



\section*{Acknowledgement}
This work has received funding from the European Union's Horizon 2020 research and innovation programme under the Marie \mbox{Sklodowska-Curie} grant agreement No 675891 \mbox{(SCAVENGE)} and by Spanish MINECO grant TEC2017-88373-R (5G-REFINE).

\bibliographystyle{IEEEtran}
\bibliography{ref}

\end{document}